\documentclass{aa}  

\usepackage{graphicx}
\usepackage[varg]{txfonts}
\begin{document} 
\titlerunning{The X--ray evolution of GRB 170817A}
\authorrunning{D'Avanzo et al.}

\title{The evolution of the X--ray afterglow emission of GW\,170817 / GRB 170817A in {\it XMM-Newton} observations}

\author{P. D'Avanzo\inst{1}, S. Campana\inst{1}, O.S. Salafia\inst{2,1,3,4}, G. Ghirlanda\inst{1,2,3}, G. Ghisellini\inst{1}, A. Melandri\inst{1}, M.G. Bernardini\inst{5,1}, M. Branchesi\inst{4,6}, E. Chassande-Mottin\inst{7}, S. Covino\inst{1}, V. D'Elia\inst{8}, L. Nava\inst{9,1}, R. Salvaterra\inst{10}, G. Tagliaferri\inst{1}, S.D. Vergani\inst{11,1}}

\institute{
INAF - Osservatorio Astronomico di Brera, Via E. Bianchi 46, I-23807, Merate (LC), Italy \and
Dipartimento di Fisica 'G. Occhialini', Universit{\`a} degli Studi di Milano-Bicocca, Piazza della Scienza 3, I-20126 Milano, Italy \and 
INFN –- Sezione di Milano-Bicocca, Piazza della Scienza 3, I-20126 Milano, Italy \and 
Gran Sasso Science Institute, Viale F. Crispi 7, L'Aquila, Italy \and
Laboratoire Univers et Particules de Montpellier, Universit{\'e} Montpellier, CNRS/IN2P3, Montpellier, France  \and
INFN, Laboratori Nazionali del Gran Sasso, I-67100, L'Aquila, Italy \and
APC, Universit{\'e} Paris Diderot, CNRS/IN2P3, CEA/Irfu, Obs de Paris, Sorbonne Paris Cit{\'e}, France \and
Space Science Data Center, ASI, Via del Politecnico, s.n.c., 00133, Roma, Italy \and 
INAF, Osservatorio Astronomico di Trieste, Via G.B. Tiepolo 11, I-34143 Trieste, Italy \and
INAF, Istituto di Astrofisica Spaziale e Fisica Cosmica di Milano, via E. Bassini 15, I-20133 Milano, Italy \and
GEPI, Observatoire de Paris, PSL Research University, CNRS, Place Jules Janssen, 92190, Meudon, France
\\
		\email{paolo.davanzo@brera.inaf.it}
                   }

   \date{}

 
\abstract{
We report our observation of the short GRB\,170817A, associated to the binary neutron star merger event GW\,170817, perfomed in the X--ray band with {\it XMM--Newton} 135 d after the event (on the 29th December 2017). We find evidence for a flattening of the X--ray light curve with respect to the previously observed brightening. This is also supported by a nearly simultaneous optical {\it Hubble Space Telescope} and successive X--ray {\it Chandra} and low-frequency radio observations recently reported in the literature. Since the optical--to--X--ray spectral slope did not change with respect to previous observations, we exclude that the change in the temporal evolution of the light curve is due to the passage of the cooling frequency: its origin must be geometric or dynamical.
We interpret all the existing afterglow data with two models:
i) a structured jet and ii) a jet--less isotropic fireball with some 
stratification in its radial velocity structure.
Both models fit the data and predict that the radio flux 
must decrease simultaneously with the optical and the X--ray one, making hard to distinguish between them at the present stage. 
Polarimetric measures and the rate of short GRB-GW association in future LIGO/Virgo runs will be key to disentangle these two geometrically different scenarios.
}
 


   \maketitle
%

\section{Introduction}
A gravitational wave (GW) event originated by the merger of a binary neutron star (BNS) system was detected for the first time by aLIGO/Virgo \citep[GW\,170817;][]{Abbott17a}, and it was found to be associated to the weak short GRB\,170817A detected by the {\it Fermi} and {\it INTEGRAL} satellites \citep{Goldstein17,Savchenko17}, marking the dawn of multi--messenger astronomy \citep{Abbott17b}. The proximity of the event \citep[$\sim 41$ Mpc;][Cantiello et al., ApJ in press]{Hjorth17} and the relative accuracy of the localization ($\sim 30$ deg$^2$, thanks to the joint LIGO and Virgo operation) led to a rapid ($\Delta t < 11$ hr) identification of a relatively bright optical electromagnetic counterpart (EM), named AT2017gfo, in the galaxy NGC\,4993 \citep{arcavi17,coulter17,lipunov17,melandri17,tanvir17,soares17,valenti17}. The analysis and modelling of the spectral characteristics of this source, together with their evolution with time, resulted in a good match with the expectations for a ``kilonova'' \citep[i.e.\ the emission due to radioactive decay of heavy nuclei produced through rapid neutron capture;][]{li98}, providing the first compelling observational evidence for the existence of such elusive transient sources \citep{cowperthwaite17,drout17,evans17,kasliwal17,nicholl17,pian17,smartt17,villar17}.
While the bright kilonova associated to GW 170817 has been widely studied and its main properties relatively well determined, the observations of the short GRB 170817A are more challenging for the current theoretical frameworks. 
Indeed, the properties of this short GRB appear puzzling in the context of observations collected over the past decades \citep{berger14,davanzo15,ghirlanda15}. The prompt $\gamma$--ray luminosity was significantly fainter (by a factor $\sim 2500$) than typical short bursts \citep[see, e.g.,][]{davanzo14}. 
A faint afterglow was detected in the X--ray and radio bands only at relatively late-times \citep[starting from $\sim$9 and 16 d after the GW/GRB trigger, respectively;][]{alexander17,haggard17,hallinan17,margutti17,troja17a}, while earlier observations provided only upper limits in these bands \citep{evans17,hallinan17}. 

Similarly to long bursts, short GRBs are thought to be produced by a relativistic jet with a typical half--opening angle $\theta_{\rm jet} \sim $5--15 deg \citep{fong16}. However, it is still unsettled whether BNS mergers can always efficiently produce a relativistic jet \citep{pasch15,ruiz16,kawamura16}. Given the small probability that our line of sight had been within the jet half--opening angle, $1-\cos(\theta_{\rm jet})$, it is unlikely that the first short GRB associated to a GW event had a jet pointing towards the Earth. 
The extremely low $\gamma$-ray luminosity of GRB 170817A has been interpreted as due to 
(i) the debeamed radiation of a jet observed off--beam (i.e. viewing angle $\theta_{\rm view} > \theta_{\rm jet}$), 
provided that the jet bulk Lorentz factor is significantly smaller than usually assumed \citep[see, e.g.,][]{pian17}. 
Alternatively, the jet could be (ii) structured, with a fast and energetic inner core surrounded by a slower, less energetic layer/sheath/cocoon (first proposed for long GRBs -- \citealt{lipunov01,rossi02,salafia15} -- and only recently extended to SGRBs -- \citealt{kathi17, Lazzati17a,gottlieb17,Lazzati17b,Lyman18,margutti18,troja18a}). In this scenario the faint, off--beam emission is due to the slower component, which originates from the interaction of the jet with the merger dynamical ejecta or the post--merger winds. 
Recently, \citet{Mooley17} 
suggested the  possibility that (iii) the jet was not successful in excavating its way through the ambient medium and that GRB\,170817A was due to its vestige, a quasi--isotropic cocoon with a velocity profile. 
Last but not least, (iv) a jet--less interpretation of GRB\,17017A could still be viable: an isotropic fireball, expanding ahead of the kilonova ejecta, which could account for both the low luminosity of the $\gamma$--ray event and the properties of the EM component in the radio and X--ray bands \citep{salafia17}. 
In this case, all BNS mergers should have this kind of faint, hard X--ray counterpart. All the above scenarios have relatively clear predictions for the temporal and spectral evolution of the electromagnetic emission from X-rays to the radio band. A comprehensive discussion of the possible physical scenarios leading to the observed broad-band emission of GW\, 170817 / GRB\,170817A can be found in Nakar \& Piran (2018).
Recent radio and X--ray observations \citep{Mooley17,margutti18,ruan2018,troja18a}, carried out until $\sim 110-115$ d after the event, indicate that the source flux is steadily rising and that the spectral energy distribution over these bands is consistent with a single power--law component. These results disfavor the interpretation (i) reported above (an off--beam homogeneous jet). 

In this letter we present deep, late--time X--ray observations of GW\, 170817 / GRB\,170817A carried out $\sim$135 days after the event with the {\it XMM-Newton} satellite, showing evidence for a a change in the temporal slope, indicating a flattening in the afterglow emission (\S 2). 
In \S 3 we interpret and discuss all the available afterglow data of GW\, 170817 / GRB\,170817A under the structured jet and jet--less scenarios mentioned above and summarize our conclusions in \S 4.

\section{Observations and data analysis}

\begin{figure}
   \centering
   \includegraphics[width=8.0cm]{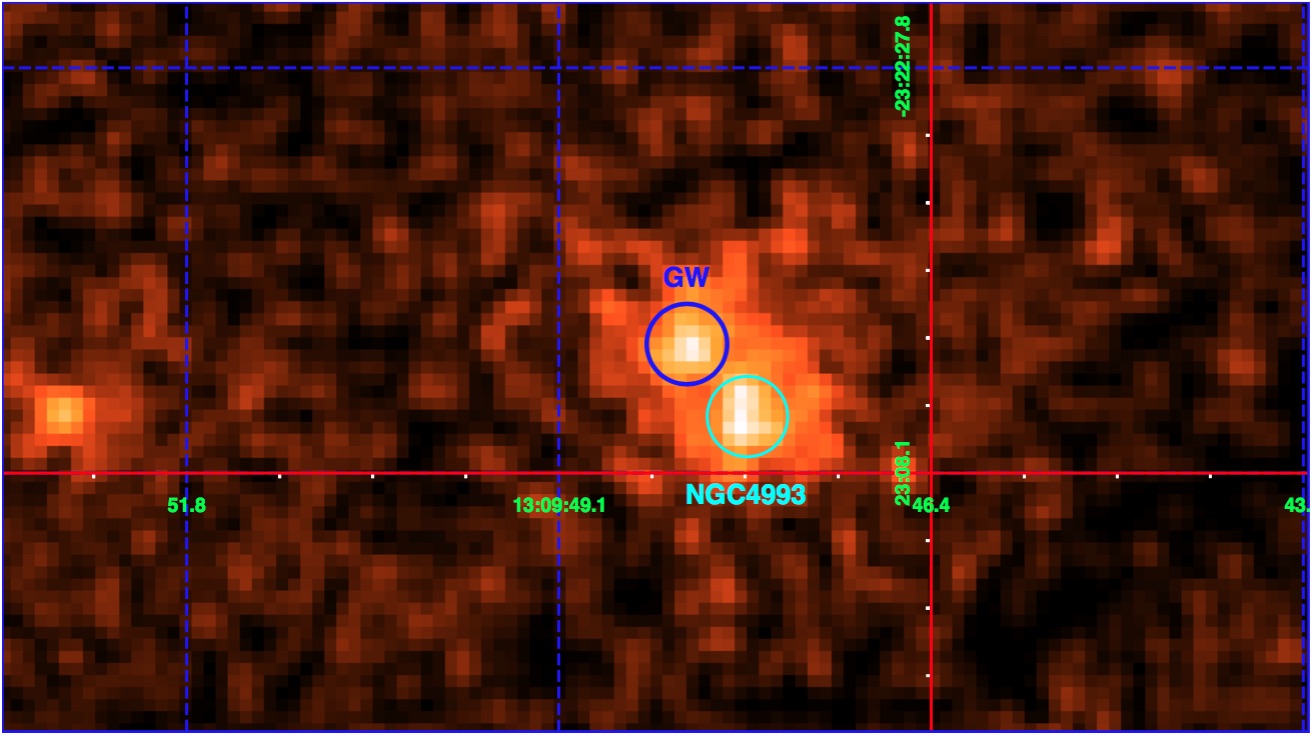}
   \caption{X--ray image obtained by co--adding the {\it XMM--Newton} pn and MOS data presented in this paper. The X--ray emission of GW\,170817 / GRB\,170817A (upper circle, 4'' radius) is clearly visible close to the nucleus of its host galaxy NGC\,4993 (lower circle).}
   \label{FigXMM}%
\end{figure}

{\it XMM--Newton} started observing GW\,170817 on the 29th December 2017 at 19:00:11 UT, 134.5 d after the GW event. 
{\it XMM--Newton} observed for 41.3 ks (42.8 ks) with the pn (MOS) detector, all equipped with the thin filter. Two large background flares occurred during the observation, reducing the usable time to 26.0 and 29.6 ks for the pn and MOS detectors, respectively.
The centre of NGC 4993 lies at only $10''$ from GW\,170817 (see Fig. \ref{FigXMM}) and particular care has to be adopted. 
We extracted products using a $4''$ radius region centered on the optical position of GW\,170817 / GRB\,170817A \citep{coulter17}. The background has been extracted from two $4''$ regions, at the same distance from the host galaxy centre, one opposite to GW\,170817 and the other to the north-east (thus avoiding the source detected by {\it Chandra}, named ``S2'' in Margutti et al. 2018, in the south--west region). We gathered (source plus background) 60, 15 and 15 counts from the pn, MOS1 and MOS2 detectors, respectively, with the source making $70-80\%$ of the total. Response matrices were generated with the package {\it SAS} v16.1 using the latest calibration products. 

Spectra were rebinned to 5 counts per spectral bin and C--statistics was adopted for the fits. We fit the three spectra with an absorbed power law model with the column density fixed to Galactic value of $7.84\times 10^{20}$ cm$^{-2}$ \citep{kalberla2005}. A $90\%$ confidence level (c.l.) upper limit on the intrinsic absorption is $<1.1\times 10^{21}$ cm$^{-2}$.
The best fit provides a photon index $\Gamma=1.7^{+0.5}_{-0.4}$ ($90\%$ c.l.) and a 0.3--10 keV unabsorbed flux ${\it F_X} = (2.1^{+1.1}_{-0.8})\times 10^{-14}$ erg cm$^{-2}$ s$^{-1}$. This fit and its uncertainty region are shown in Fig. \ref{SPXMM} by the dot--dashed line and the grey shaded region, respectively. The {\it XMM--Newton} de--absorbed data are also shown in Fig. \ref{SPXMM} (blue points). 

Motivated by the almost simultaneous {\it XMM--Newton} and {\it HST} observations ($\sim$137 d after the event; \citealt{margutti18}), we dereddened\footnote{We corrected for Galactic extinction assuming $E(B-V) = 0.105$ mag \citep{schlafly11}.} the optical AB magnitude mag$_{F606W}=26.90\pm0.25$ reported by these authors and we fitted together optical and X--ray data. 
Thanks to the large leverage in terms of energy range we tightly constrain the overall power law photon index to $\Gamma=1.60\pm0.05$. This fit is shown by the dashed red line and its uncertainty by the yellow shaded region in Fig. \ref{SPXMM}.
Adopting this index in the 0.3--10 keV, the unabsorbed flux is $F_X = (2.1^{+0.7}_{-0.5})\times 10^{-14}$ erg cm$^{-2}$ s$^{-1}$, which translates to a luminosity $L_X = 4\times10^{39}$ erg s$^{-1}$ (at 41 Mpc).
The {\it XMM-Newton} flux and photon index are fully consistent with the values found about one month before and after our observation by {\it Chandra} (namely, a 0.3--10 keV unabsorbed flux $F_X = (2.5 \pm 0.3)\times 10^{-14}$  erg cm$^{-2}$ s$^{-1}$ and $F_X = (2.6 \pm 0.3)\times 10^{-14}$ erg cm$^{-2}$ s$^{-1}$ measured at $\sim 109$ and $159$ days after the GW trigger, respectively; \citealt{margutti18,troja18a,troja18b}) and indicate that the GW\,170817 flux stopped increasing. 

\section{Interpretation and discussion}

\begin{figure}
\centering
\includegraphics[width=\columnwidth]{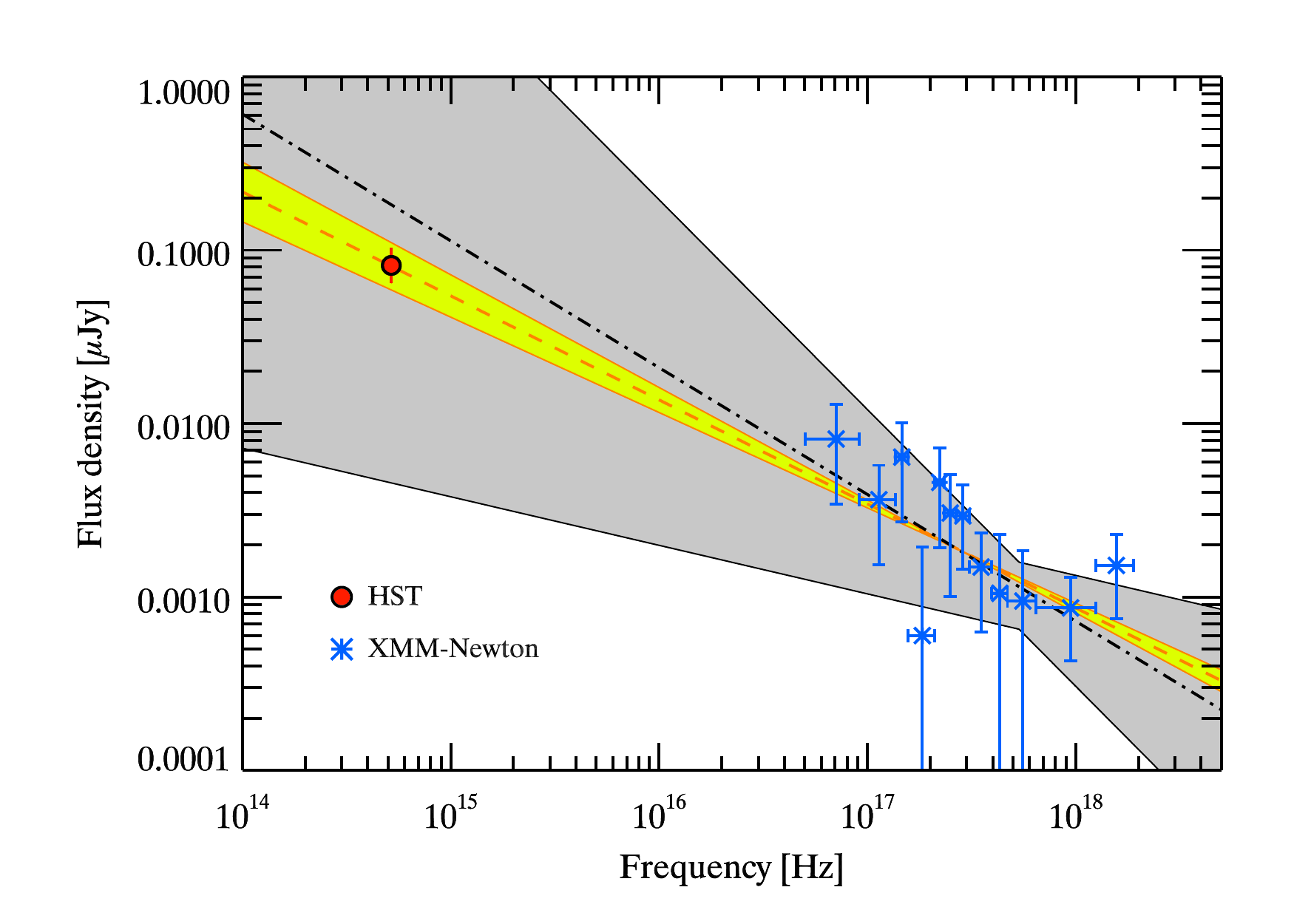}
\caption{
Optical to X--ray spectrum of GRB 170817A. 
{\it XMM--Newton} data points (blue, {\it this work}) and {\it HST} 
contemporaneous detection (red circle) are shown.
The gray shaded region shows the 90\% uncertainty on the fit of 
the {\it XMM--Newton} data alone (dot--dashed line). 
The fit obtained combining the de--reddened HST flux 
(from Margutti et al. 2018) and our {\it XMM-Newton} data results in a  
photon index of 1.6 (dotted red line) with an error of $\pm 0.05$ 
(90\% c.l. -- yellow shaded region).}
\label{SPXMM}
\end{figure}

Fig.~\ref{fig:lightcurves} shows the afterglow data published in the literature to date, together with our {\it XMM--Newton} point obtained at $\sim 135$ d (light blue circle). All radio and X--ray detections of GRB 170817A reported so far covering the time range between $\sim$9 and $\sim$115 d after the event indicated a steady increase of the source flux ($F(t)\propto t^{0.7-0.8}$), with negligible spectral evolution over a broadband spectrum ($F_{\nu} \propto \nu^{-0.6}$; \citealt{Mooley17,margutti18,ruan2018,troja18a}). By comparing the nearly simultaneous {\it XMM--Newton} and {\it HST} observations (see Sect. 2) we find that the spectral energy distribution (SED) at the epoch of our {\it XMM--Newton} observation is still consistent with a single power--law component between the X--ray and the optical, with $F_{\nu}\propto \nu^{-0.6}$ (Fig. \ref{SPXMM}). 
Given the lack of spectral evolution, we found reasonable to assume that the light curve is evolving in the same way at all wavelengths and carried out a a combined fit of the available radio (3 GHz), optical\footnote{Concerning the optical band, we included in our fit only the {\it HST} data ({\it F606W} filter) obtained at $\sim$110 and $\sim$137 d after the GW trigger, i.e., those obtained when the thermal component due to the kilonova emission is not contributing anymore \citep{Lyman18,margutti18}.} and X--ray (0.3-10 keV) data obtained between $\sim$9 and $\sim$159 d after the event  \citep{hallinan17,Mooley17,Lyman18,margutti18,troja18a,troja18b}. Using an F-test we compared the joined fit obtained with a single and a broken power--law model. We find that a temporal break is required at $1.94\sigma$ (95\% c.l.). While with a single power-law ($F(t)\propto t^{\alpha}$) model we obtain an index $\alpha \sim 0.8$, in agreement with other findings \citep{Mooley17,margutti18,troja18a}, at the same time we note that both the {\it XMM-Newton} (this work) and {\it Chandra}\footnote{A different result, with a X-ray light curve consistent with a $t^{0.7}$ rise, is found by Haggard et al. (2018), based on the same {\it Chandra} observations used by \citet{troja18b}, although these authors reported their results in a sligthly different energy band.} \citep{troja18b} data points obtained at successive epochs over two months fall below the extrapolation of the best fit (by 1.5 and 2.7 sigma, respectively). Such an increasing divergence, together with the mild indication of a temporal break provided by the F-test, is indicative of a change in the slope, with a flattening, of the X--ray light curve with respect to the brightening trend observed in the X--ray and radio bands at earlier epochs. This change in the light curve temporal evolution is observed also in the optical band by the {\it HST} observations carried out at $\sim 110$ and $\sim 137$ d with {\it HST} (\citealt{Lyman18,margutti18}; see also Fig.~3). Furthermore, the evidence for a possible plateau in the light curve has been recently found in low-frequency radio data obtained between 66 and 152 days after the event (Resmi et al. 2018), although the relatively poor temporal sampling prevents to firmly exclude a rising trend. Overall, the multi-wavelength behaviour of the afterglow provides an indication that the light curve is changing slope (see also \citealt{Dobie18}), even if it is too early to constrain the temporal evolution after the break, since at this epoch we are still sampling its turning point.

The X--ray spectrum expected if the cooling frequency has transitioned below the X--ray band (between the earlier {\it Chandra} observation and our {\it XMM-Newton} epoch) has a photon index of 2.1 and predicts an optical flux which is inconsistent with that observed by {\it HST} nearly at the same epoch. We can thus exclude this explanation of the observed optical and X--ray light curve peak. We rather interpret it as due to a dynamical or geometric effect. 
We interpret the optical and X--ray light curve within two possible scenarios: 
(i) a structured jet, in which case the decline of the optical and X--ray fluxes indicates that the emission from the jet core has entered our line of sight (i.e. the core has decelerated down to a Lorentz factor $\Gamma\sim \theta_\mathrm{view}^{-1}$);
(ii) an isotropic (Salafia et al. 2017) and stratified fireball with a velocity profile, as that described in \citet{Mooley17}, in which case the light curve peak indicates that the velocity profile has a rather sharp cut--off at $\beta_\mathrm{min} = v_\mathrm{min}/c\sim 0.88$. 


\subsection{The structured jet model}

\begin{figure*}
\centering
\includegraphics[width=\textwidth]{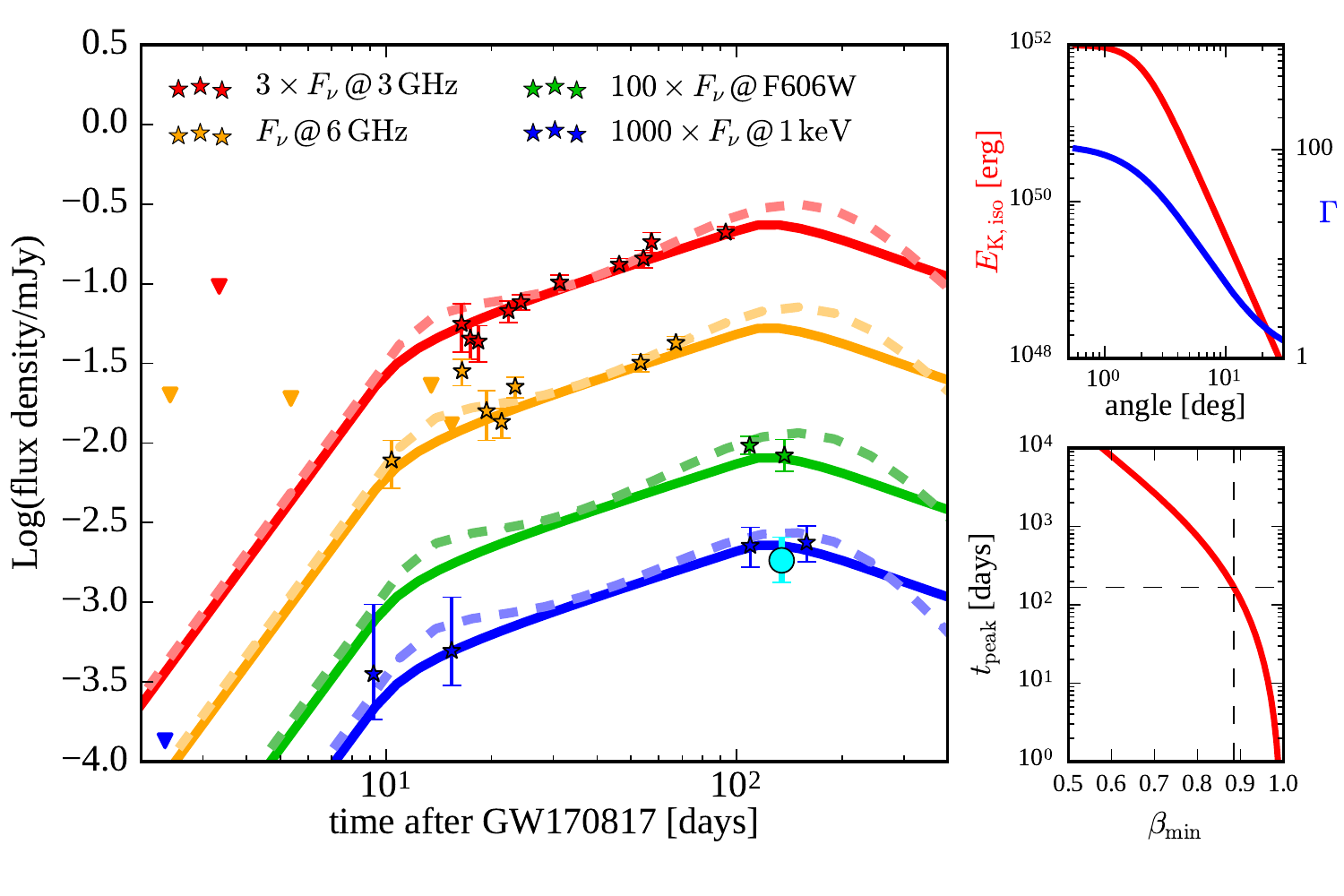}  
\vskip -0.5 cm
\caption{\textbf{Left--hand panel:}
GRB 170817A afterglow light curves in radio at 3 GHz and 6 GHz (red and orange 
stars respectively, VLA observations -- data from \citealt{hallinan17,Mooley17}), in the 
optical (green stars, HST/ACS observations in the F606W filter -- data from \citealt{Lyman18} and \citealt{margutti18}) and in the X--rays (blue stars: {\it Chandra} observations, data from \citealt{margutti18,troja18b}; light blue circle: our \textit{XMM--Newton} observation). 
Thick coloured solid lines represent our isotropic fireball model (corresponding to either the jet--less scenario outlined in \citealt{salafia17}, or the choked jet scenario proposed by \citealt{Mooley17}). The brown dashed lines represent our structured jet model. The parameters of both models are reported in Table~\ref{tab:sj_params}. \textbf{Upper right--hand panel:} the jet structure assumed in our model. The red line represents the isotropic equivalent kinetic energy, while the blue line shows the Lorentz factor. \textbf{Lower right--hand panel:} the red line shows the peak time of the isotropic outflow light curve as a function of the minimum velocity $\beta_\mathrm{min}$ in the velocity profile. The dashed lines mark the value we employed in the modeling.
}
\label{fig:lightcurves}
\end{figure*}

If a jet is launched after the merger, it must excavate its way out of the inner region, which can be baryon-polluted due to the post-merger winds and the dynamical ejecta. The propagation through such ambient material is likely to have a major role in shaping the jet angular distribution of energy and terminal Lorentz factor at breakout (see e.g. the simulations by Lazzati et al. 2017b). The resulting jet structure features an inner, narrow, faster core with a relatively uniform distribution of kinetic energy per unit solid angle, surrounded by a slower, extended structure whose kinetic energy per unit solid angle decreases relatively fast with the distance from the jet axis. This latter structure can be identified as the vestige of the jet cocoon (constituted by the jet and ambient material that has been shocked during the excavation). Guided by this picture, we employ a simple structured jet model, in which both the isotropic 
equivalent kinetic energy $E_\mathrm{K,iso}$ and the bulk Lorentz 
factor $\Gamma$ are approximately constant within a narrow core of 
half--opening angle $\theta_\mathrm{core}$ and decrease as power--laws outside of it:
\begin{equation}
E_\mathrm{K,iso}(\theta) = \frac{E_\mathrm{K,iso,core}}{1+\left(\theta/\theta_\mathrm{core}\right)^{s_1}}
\end{equation}
and
\begin{equation}
\Gamma_0(\theta) = 1 + \frac{\Gamma_\mathrm{core} - 1}{1+\left(\theta/\theta_\mathrm{core}\right)^{s_2}}
\end{equation}
We model the dynamics of the jet with the simplifying assumption that each 
solid angle element evolves independently (i.e.\ we neglect side expansion, 
which should have a limited effect on the light curve, see e.g. \citealt{Granot03,Lazzati17b,Lamb17}). 
For each solid angle element, we model the emission following \citet{Sari98}, with the proper 
transformations to the off--axis observer frame. 
The ambient medium is assumed to have a constant number density $n$. 
The parameters of the structured jet model shown in Fig.~\ref{fig:lightcurves} (dashed lines) are reported in Table~\ref{tab:sj_params} (where $\epsilon_e$ and $\epsilon_B$ are the shock energy carried by the electrons and by the magnetic field, respectively, and $p$ is the electron energy distribution index). With the given parameters, the total kinetic energy in the jet is $E_\mathrm{jet}\approx 1.1\times 10^{49}\,\mathrm{erg}$ (for one jet), which is just what is expected for a standard SGRB jet \citep{Hotokezaka16}. 
Different structured jet scenarios have been proposed to model the afterglow light curves of GRB\,170817A by \cite{Lazzati17b}, \cite{Lyman18}, \cite{margutti18} and \cite{troja18a}. As in our case described above, the models presented in \cite{Lyman18}, \cite{margutti18} and \cite{troja18a} can account for the change in the slope observed in the X-ray and optical light curve at $t \sim 110-130$ d, predicting a relatively long plateau at these epochs.
We note, however, that all the proposed model are very similar and that the diversity in the predictions can be ascribed to a combination of differences in the jet structure (including the opening angle of the relativistic core), the density of the environment and by the different choice of the microphysical parameters, that can be better constrained with future multi-band observations.

\noindent

\begin{table}
\caption{\label{tab:sj_params}Parameters of the structured jet and isotropic outflow models shown 
in Fig.~\ref{fig:lightcurves}. Units in square parentheses.}
\centering
\begin{tabular}{lr|lr}
\hline
\hline
\multicolumn{2}{c}{Structured Jet} & \multicolumn{2}{c}{Isotropic outflow}\\
\hline


$\theta_\mathrm{core}$ [deg] & $2$ & & \\
$E_\mathrm{K,iso,core}$ [erg] & $ 1\times 10^{52}$ & $E_0$ [erg] & $3\times 10^{51}$\\
$s_1 $ & $ 3.5$ & $\alpha$ & 5\\
$\Gamma_\mathrm{core} $ & $ 110$ & $\Gamma_\mathrm{max}$ & $3.8$ \\
$s_2   $ & $ 2$ & $\beta_\mathrm{min}$ & $0.875$\\
$\theta_\mathrm{view}$ & $22.5$ & & \\
$n $ [cm$^{-3}$] & $ 10^{-3}$ & $n $ [cm$^{-3}$] & $ 2\times 10^{-4}$\\
$\epsilon_{\rm e} $ & $ 0.06$ & $\epsilon_{\rm e} $ & $ 0.1$\\
$\epsilon_{\rm B} $ & $ 0.01$ & $\epsilon_{\rm B} $ & $ 0.01$\\
$ p$ & $2.13$ & $ p$ & $2.14$ \\

\hline
\end{tabular}
\end{table}

\subsection{The isotropic outflow model}

\citet{salafia18} proposed a scenario where a reconnection--powered isotropic fireball is launched at the beginning of the neutron--star merger phase. The simple model sketched there assumed a uniform energy profile in the fireball, however the described process may also produce a fireball with an energy profile as that described in \citet{Mooley17}. 
In what follows, we adopt a similar model as that in Mooley et al. (2017) to describe the light
curve in our jet–less scenario, with the difference that we take into account the proper equal-arrival-time
surfaces in the computation of the observed flux. In this scenario, an isotropic (or quasi–isotropic) outflow
is launched, with a distribution of energy in momentum space given by:
\begin{equation}
E(>\Gamma\beta) = E_0 (\Gamma\beta)^{-\alpha}
\end{equation}
between the minimum and maximum Lorentz factors $\Gamma_\mathrm{min}$, $\Gamma_\mathrm{max}$ or equivalently the minimum and maximum velocities $\beta_\mathrm{min}$, $\beta_\mathrm{max}$. The interaction with the ISM results in a shock whose dynamics reflect the fact that slower (but more energetic) ejecta progressively cross the reverse shock, reducing the deceleration. The evolution of the forward shock radius \citep{Hotokezaka16} is given by 
\begin{equation}
\frac{4}{3}\pi R^3 m_p n (c\beta\Gamma)^2 = E(>\Gamma\beta)
\end{equation}
where $m_p$ is the roton mass. As soon as all the outflow material has gone across the reverse shock, i.e.\ after the minimum ejecta velocity 
$\beta_\mathrm{min}$ has been reached, the dynamics turn into simple adiabatic expansion, 
with $\Gamma\beta \propto R^{-3/2}$ \citep{Nava13}. We model the synchrotron emission from the shock-heated electrons following \citet{Sari98}, just as in the structured jet model. The parameters are essentially the same as in \citet{Mooley17}, except for the slightly lower value of the electron energy power law slope $p=2.14$, which provides a better agreement to the broadband spectrum (see e.g. \citealt{margutti18} therein Fig.~6), and for the introduction of the minimum ejecta velocity 
$\beta_\mathrm{min}$ in order to account for the peak in the light curve. The parameters of the isotropic outflow model shown in Fig.~\ref{fig:lightcurves} (solid lines) are reported in Table~\ref{tab:sj_params}. The value $\beta_\mathrm{min}=0.875$ we employed in the modeling implies a total kinetic energy $ E_\mathrm{tot}\approx 1.6 \times 10^{50}\,\mathrm{erg}$ (assuming spherical geometry).

\section{Conclusions}

The {\it XMM--Newton} late time observations of the afterglow of GRB\,170817A associated to the BNS merger event GW\,170817 presented in this work show evidence that the X--ray flux has flattened during the last two months (Dec 2017 - Jan 2018). This is supported by the latest {\it HST} observations at the same epoch \citep{margutti18}, by later {\it Chandra} X-ray observations \citep{troja18b} {and by late-time GMRT low-frequency radio observations (Resmi et al. 2018)}. The combined spectrum obtained with nearly-simultaneous {\it XMM-Newton} and {\it HST} data show no spectral evolution with respect to previous observations, suggesting  a geometric or dynamical origin for the decrease in flux observed in the afterglow light curve. We modelled the observed X-rays, optical and radio afterglow emission as (i) the deceleration peak of the core of a structured jet (as described in \S 3.1) pointing away from our line of sight or (ii) the deceleration of an isotropic fireball with a radial velocity structure. We found that both models succesfully reproduce the available data, that is not surprising since in both cases we are still observing the emission from the slower ejecta. 
The similarity of the light curves of the two models as shown in Fig. \ref{fig:lightcurves} may require some independent measure to disentangle between these two possible scenarios. 
A possible diagnostic test able to discriminate between isotropic and jetted
geometries is based on linear polarisation measurements given that, to observe polarization, some degree of asymmetry is needed (Rossi et al. 2004; Covino \& Gotz 2016 and references therein). A
general prediction for late-time afterglows is that they can show linear
polarization even up to at a fairly high level ($\sim$10\%) depending on the assumed geometry (see
\citealt{rossi2004} for more details). With
the presently available technologies, for the AT2017gfo at late times,  such a
measurement is very demanding and could only be feasible (in the most optimistic case, i.e. for polarization level of $5-10$\%) at radio wavelengths\footnote{Optical polarization studies should in principle be discriminant as well, but the source is too faint in this band.}. On the contrary, no polarization is essentially
expected for an isotropic emission. However, while detection of linear polarization
would be a clear indication of a jetted geometry, a null result may not be
conclusive. At radio frequencies linear polarization can be detected at frequencies
higher than the self--absorption frequency but it can also be suppressed by Faraday
rotation depending on the specific micro-physical parameters (Toma et al. 2008),
making the interpretation of null polarisation, without meaningful observations at
higher frequencies, hard and possibly inconclusive\footnote{Further discussion and predictions about radio polarimetry of GW\,170817 have been recently reported by Geng et al. (2018) and Gill \& Granot (2018).}.
Besides this, such a different geometry is expected to significantly affect the rate of burst similar to GRB\,170817A observed in association to GW events detected during the forthcoming LIGO/Virgo observing runs (Ghirlanda et al., in prep.).


\begin{acknowledgements}
We thank Norbert Schartel and the {\it XMM-Newton} staff for approving, scheduling and carrying out these observations. We thank A. Possenti for useful discussion. We acknowledge support from ASI grant I/004/11/3. MGB acknowledges support of the OCEVU Labex (ANR-11-LABX-0060) and the A*MIDEX project (ANR-11-IDEX-0001-02) funded by the ``Investissements d'Avenir'' French government program managed by the ANR.
\end{acknowledgements}


\end{document}